%% file: main.tex
\documentclass{article}
\usepackage[OT1]{fontenc}
\usepackage{spconf,amsmath,graphicx,hyperref}
\usepackage{amssymb,booktabs,xcolor}
\hypersetup{hidelinks}

\newcommand{\SetResult}[2]{\expandafter\def\csname resultvalue:#1\endcsname{#2}}
\newcommand{\Result}[1]{%
  \ifcsname resultvalue:#1\endcsname
    \csname resultvalue:#1\endcsname
  \else
    \mbox{\textemdash}%
  \fi}
\newcommand{\Speech}{\textsc{Speech}}
\newcommand{\Music}{\textsc{Music}}
\newcommand{\Sound}{\textsc{Sound}}
\newcommand{\Best}[1]{{\bfseries #1}}
\newcommand{\TableSize}{\footnotesize}
\newcommand{\TableBodySize}{\footnotesize}
\input{result_values}
\input{derived_values}

\title{Semantic Refinement of Universal Audio Representations through Audio-Description Alignment}
\name{\shortstack{Lejun Min\textsuperscript{1,2}, Junyu Dai\textsuperscript{1},
Ruichen Zheng\textsuperscript{1}, Xinyue Fan\textsuperscript{1},
Yang Xiang\textsuperscript{1}, Huaichen Zhang\textsuperscript{1}\\
Xingchen Song\textsuperscript{1}, Yufei Shi\textsuperscript{1},
Han Zhao\textsuperscript{1}, Xiangang Li\textsuperscript{1}}}
\address{\textsuperscript{1}Alibaba Token Foundry\\
\textsuperscript{2}Center for Computer Research in Music and Acoustics, Stanford University}

\begin{document}
\ninept
\maketitle

\begin{abstract}
Universal audio representations must preserve acoustic detail while making high-level
concepts accessible across speech, music, environmental sound, and downstream models
of different capacities. We study semantic refinement of an acoustically pretrained
encoder by adding audio--description alignment to a foundation of BEST-RQ,
reconstruction, and CTC. We compare matched control, shuffled-description, and
correctly paired trajectories to distinguish correct correspondence from an extra
contrastive objective. Each endpoint is frozen and evaluated with a temporal-mean linear probe and
a sequence-aware LLM readout, testing whether the refined information is directly
accessible and remains useful to a stronger model.
Across three paired seeds, correct alignment improves domain-balanced classification
by \Result{linearcontrast.p3-p1.equal_domain.mean} points with the linear probe and
\Result{contrast.p3-p1.equal_domain.mean} points with the sequence-aware LLM, with
positive changes in every domain. Correct pairing accounts for 87\% of the linear-probe
gain, while the LLM shows its clearest correspondence-specific benefit in captioning.
Dense acoustic objectives provide complementary gains under both readouts. A separate
24-layer continuation remains competitive with leading public encoders under the
shared evaluator, supporting the recipe beyond the controlled study.
\end{abstract}


\input{sections/introduction}
\input{sections/method}
\input{sections/experimental_design}
\input{sections/results}
\input{sections/conclusion}
\clearpage

\bibliographystyle{IEEEbib}
\bibliography{references}
\end{document}

%% file: result_values.tex
\SetResult{proxy.p0rq.asr.mean}{45.30}
\SetResult{proxy.p1.equal_domain.mean}{67.88}
\SetResult{proxy.p1.speech.mean}{78.00}
\SetResult{proxy.p1.music.mean}{68.28}
\SetResult{proxy.p1.sound.mean}{57.38}
\SetResult{proxy.p3.equal_domain.mean}{70.48}
\SetResult{proxy.p3.speech.mean}{79.12}
\SetResult{proxy.p3.music.mean}{71.60}
\SetResult{proxy.p3.sound.mean}{60.71}
\SetResult{contrast.p3-p1.equal_domain.mean}{2.59}
\SetResult{contrast.p3-p1.equal_domain.ci-low}{1.42}
\SetResult{contrast.p3-p1.equal_domain.ci-high}{3.77}
\SetResult{contrast.p3-p1.speech.mean}{1.13}
\SetResult{contrast.p3-p1.music.mean}{3.32}
\SetResult{contrast.p3-p1.sound.mean}{3.33}
\SetResult{contrast.p3-p1.task2.mean}{2.09}
\SetResult{contrast.p3-p1.asr.mean}{0.67}
\SetResult{contrast.p3-p1.caption.mean}{3.03}
\SetResult{contrast.p5-p2.equal_domain.mean}{0.11}
\SetResult{contrast.p5-p2.asr.mean}{-0.15}
\SetResult{contrast.p5-p2.caption.mean}{0.14}
\SetResult{contrast.p5-p3.equal_domain.mean}{0.78}
\SetResult{contrast.p5-p3.asr.mean}{-0.18}
\SetResult{contrast.p5-p3.caption.mean}{0.11}
\SetResult{linear.p1.equal_domain.mean}{63.55}
\SetResult{linear.p1.speech.mean}{74.11}
\SetResult{linear.p1.music.mean}{62.30}
\SetResult{linear.p1.sound.mean}{54.24}
\SetResult{linear.p3.equal_domain.mean}{68.20}
\SetResult{linear.p3.speech.mean}{75.16}
\SetResult{linear.p3.music.mean}{68.05}
\SetResult{linear.p3.sound.mean}{61.39}
\SetResult{linearcontrast.p3-p1.equal_domain.mean}{4.66}
\SetResult{linearcontrast.p3-p1.equal_domain.ci-low}{4.43}
\SetResult{linearcontrast.p3-p1.equal_domain.ci-high}{4.89}
\SetResult{linearcontrast.p3-p1.speech.mean}{1.06}
\SetResult{linearcontrast.p3-p1.music.mean}{5.76}
\SetResult{linearcontrast.p3-p1.sound.mean}{7.16}
\SetResult{linearcontrast.p3-p1.task.cremad.mean}{1.32}
\SetResult{linearcontrast.p3-p1.task.fsc.mean}{-0.44}
\SetResult{linearcontrast.p3-p1.task.libricount.mean}{0.47}
\SetResult{linearcontrast.p3-p1.task.scv1.mean}{0.67}
\SetResult{linearcontrast.p3-p1.task.vocalsound.mean}{2.28}
\SetResult{linearcontrast.p3-p1.task.voxceleb1.mean}{3.72}
\SetResult{linearcontrast.p3-p1.task.voxlingua33.mean}{-0.62}
\SetResult{linearcontrast.p3-p1.task.fma.mean}{4.58}
\SetResult{linearcontrast.p3-p1.task.gtzan.mean}{8.42}
\SetResult{linearcontrast.p3-p1.task.nsynth.mean}{4.27}
\SetResult{linearcontrast.p3-p1.task.esc50.mean}{11.50}
\SetResult{linearcontrast.p3-p1.task.fsd50k.mean}{6.88}
\SetResult{linearcontrast.p3-p1.task.fsdkaggle2018.mean}{4.94}
\SetResult{linearcontrast.p3-p1.task.urbansound8k.mean}{5.30}
\SetResult{linearfactorial.none.equal_domain}{62.55}
\SetResult{linearfactorial.none.music}{61.11}
\SetResult{linearfactorial.none.sound}{53.43}
\SetResult{linearfactorial.music.equal_domain}{65.78}
\SetResult{linearfactorial.music.music}{67.77}
\SetResult{linearfactorial.music.sound}{55.48}
\SetResult{linearfactorial.sound.equal_domain}{65.91}
\SetResult{linearfactorial.sound.music}{63.80}
\SetResult{linearfactorial.sound.sound}{59.78}
\SetResult{linearfactorial.both.equal_domain}{67.54}
\SetResult{linearfactorial.both.music}{66.61}
\SetResult{linearfactorial.both.sound}{61.36}
\SetResult{linear.full.domainsemantic105k.equal_domain}{73.59}
\SetResult{factorial.none.music}{68.60}
\SetResult{factorial.none.sound}{57.25}
\SetResult{factorial.none.equal_domain}{67.88}
\SetResult{factorial.music.music}{71.50}
\SetResult{factorial.music.sound}{57.38}
\SetResult{factorial.music.equal_domain}{68.68}
\SetResult{factorial.sound.music}{67.83}
\SetResult{factorial.sound.sound}{59.98}
\SetResult{factorial.sound.equal_domain}{68.77}
\SetResult{factorial.both.equal_domain}{69.79}
\SetResult{factorial.both.music}{70.97}
\SetResult{factorial.both.sound}{59.77}
\SetResult{xares.public.wavlm.asr}{79.55}
\SetResult{xares.public.wavlm.caption}{50.50}
\SetResult{xares.public.wavlm.equal_domain}{68.67}
\SetResult{xares.public.muq.asr}{25.00}
\SetResult{xares.public.muq.caption}{52.53}
\SetResult{xares.public.muq.equal_domain}{71.03}
\SetResult{xares.public.dasheng.asr}{70.45}
\SetResult{xares.public.dasheng.caption}{54.77}
\SetResult{xares.public.dasheng.equal_domain}{77.43}
\SetResult{xares.public.spear.asr}{82.80}
\SetResult{xares.public.spear.caption}{55.73}
\SetResult{xares.public.spear.equal_domain}{76.70}
\SetResult{xares.full.domainsemantic105k.asr}{87.45}
\SetResult{xares.full.domainsemantic105k.caption}{54.40}
\SetResult{xares.full.domainsemantic105k.equal_domain}{75.78}

%% file: derived_values.tex
\SetResult{contrast.p3-p1.task.cremad.mean}{1.67}
\SetResult{contrast.p3-p1.task.fsc.mean}{0.20}
\SetResult{contrast.p3-p1.task.libricount.mean}{2.43}
\SetResult{contrast.p3-p1.task.scv1.mean}{0.30}
\SetResult{contrast.p3-p1.task.vocalsound.mean}{0.30}
\SetResult{contrast.p3-p1.task.voxceleb1.mean}{1.97}
\SetResult{contrast.p3-p1.task.voxlingua33.mean}{1.03}
\SetResult{contrast.p3-p1.task.fma.mean}{2.20}
\SetResult{contrast.p3-p1.task.gtzan.mean}{9.43}
\SetResult{contrast.p3-p1.task.nsynth.mean}{-1.67}
\SetResult{contrast.p3-p1.task.esc50.mean}{5.13}
\SetResult{contrast.p3-p1.task.fsd50k.mean}{1.77}
\SetResult{contrast.p3-p1.task.fsdkaggle2018.mean}{2.30}
\SetResult{contrast.p3-p1.task.urbansound8k.mean}{4.13}

\SetResult{linearcontrast.p5-p3.equal_domain.mean}{1.31}
\SetResult{linearcontrast.p5-p3.equal_domain.ci-low}{0.82}
\SetResult{linearcontrast.p5-p3.equal_domain.ci-high}{1.79}
\SetResult{linearcontrast.p5-p2.equal_domain.mean}{-0.21}

\SetResult{shuffle.linear.equal_domain.mean}{64.14}
\SetResult{shuffle.linear.speech.mean}{74.02}
\SetResult{shuffle.linear.music.mean}{63.74}
\SetResult{shuffle.linear.sound.mean}{54.64}
\SetResult{shuffle.linear-minus-control.equal_domain.mean}{0.59}
\SetResult{shuffle.linear.correct-minus-shuffle.equal_domain.mean}{4.07}
\SetResult{shuffle.linear.correct-minus-shuffle.equal_domain.ci-low}{3.63}
\SetResult{shuffle.linear.correct-minus-shuffle.equal_domain.ci-high}{4.50}
\SetResult{shuffle.linear.correct-minus-shuffle.speech.mean}{1.14}
\SetResult{shuffle.linear.correct-minus-shuffle.music.mean}{4.31}
\SetResult{shuffle.linear.correct-minus-shuffle.sound.mean}{6.75}

\SetResult{shuffle.llm.shuffled.equal_domain.mean}{68.38}
\SetResult{shuffle.llm.shuffled.speech.mean}{78.32}
\SetResult{shuffle.llm.shuffled.music.mean}{68.32}
\SetResult{shuffle.llm.shuffled.sound.mean}{58.48}
\SetResult{shuffle.llm.correct-minus-shuffle.speech.mean}{0.80}
\SetResult{shuffle.llm.correct-minus-shuffle.music.mean}{3.28}
\SetResult{shuffle.llm.correct-minus-shuffle.sound.mean}{2.23}
\SetResult{shuffle.llm.correct-minus-shuffle.equal_domain.mean}{2.10}
\SetResult{shuffle.llm.correct-minus-shuffle.equal_domain.ci-low}{-1.78}
\SetResult{shuffle.llm.correct-minus-shuffle.equal_domain.ci-high}{5.98}
\SetResult{shuffle.llm.control.task2.mean}{65.13}
\SetResult{shuffle.llm.shuffled.task2.mean}{65.87}
\SetResult{shuffle.llm.correct.task2.mean}{67.22}
\SetResult{shuffle.llm.control.asr.mean}{87.33}
\SetResult{shuffle.llm.shuffled.asr.mean}{87.67}
\SetResult{shuffle.llm.correct.asr.mean}{88.00}
\SetResult{shuffle.llm.correct-minus-shuffle.asr.mean}{0.33}
\SetResult{shuffle.llm.control.caption.mean}{50.33}
\SetResult{shuffle.llm.shuffled.caption.mean}{51.34}
\SetResult{shuffle.llm.correct.caption.mean}{53.37}
\SetResult{shuffle.llm.correct-minus-shuffle.caption.mean}{2.02}
\SetResult{shuffle.llm.correct-minus-shuffle.task2.mean}{1.35}

\SetResult{p5-p4.linear.equal_domain.mean}{1.80}
\SetResult{p5-p4.linear.equal_domain.ci-low}{0.59}
\SetResult{p5-p4.linear.equal_domain.ci-high}{3.01}
\SetResult{p5-p4.llm.equal_domain.mean}{2.02}
\SetResult{p5-p4.llm.equal_domain.ci-low}{0.79}
\SetResult{p5-p4.llm.equal_domain.ci-high}{3.25}
\SetResult{p5-p4.linear.sound.mean}{4.16}
\SetResult{p5-p4.llm.sound.mean}{4.67}
\SetResult{p5-p4.llm.asr.mean}{-0.30}
\SetResult{p5-p4.llm.caption.mean}{0.93}

\SetResult{linear.p1seed1342-minus-p0rq.equal_domain}{3.24}
\SetResult{linear.p1seed1342-minus-p0rq.speech}{13.75}
\SetResult{linear.p1seed1342-minus-p0rq.music}{-2.23}
\SetResult{linear.p1seed1342-minus-p0rq.sound}{-1.77}
\SetResult{proxy.p1seed1342.asr}{86.95}
\SetResult{proxy.p1seed1342-minus-p0rq.equal_domain}{1.28}
\SetResult{proxy.p1seed1342-minus-p0rq.asr}{41.65}
\SetResult{proxy.p1seed1342-minus-p0rq.caption}{-0.03}

\SetResult{linear.public.wavlm.equal_domain}{67.75}
\SetResult{linear.public.muq.equal_domain}{66.05}
\SetResult{linear.public.dasheng.equal_domain}{76.81}
\SetResult{linear.public.spear.equal_domain}{78.18}

%% file: sections/introduction.tex
\section{Introduction}
\label{sec:intro}

As audio systems expand from domain-specific recognition toward language-conditioned
understanding and generation across speech, music, and environmental sound, a universal
representation can provide a shared interface between waveforms and downstream models.
It must preserve fine temporal and spectral structure while organizing sound around
concepts that transfer across tasks and domains. Acoustic pretraining supplies much of
the former; we use \emph{semantic refinement} for the deliberate addition of the latter.
This goal is concrete in semantic-aware music tokenizers \cite{lin2026duotok}. Dai et al.
\cite{dai2026fullsong}, for example, place one before hierarchical autoregressive
planning and flow-matching rendering. Their training path combines BEST-RQ pretraining,
ASR/reconstruction/chroma refinement, and discrete-token training: BEST-RQ learns
contextual acoustics, reconstruction retains spectral detail, chroma carries pitch, and
ASR supplies lexical supervision. These objectives form a strong acoustic--lexical
basis, yet do not explicitly organize it around scene, style, or instrumentation.

Two lines of work motivate this refinement. Masked prediction provides the acoustic
foundation \cite{chiu2022bestrq,chen2022wavlm,li2024mert,zhu2025muq,chen2023beats,
dinkel2024dasheng,pepino2024encodecmae}; audio--language methods add semantics through
clip alignment \cite{elizalde2023clap}, joint masked/contrastive or captioning objectives
\cite{niizumi2025m2d2,mei2026slap}, multidomain captions \cite{tseng2026captionstew},
and frame-level grounding \cite{li2026finelap}. Their gains establish the value of
language supervision, but are measured across different recipes, mixtures, and
downstream systems. They therefore leave unclear how much correct correspondence adds
within a matched continuation, and whether the change is directly accessible from the
representation or mainly exposed by a capable downstream model.

We address this gap by extending the encoder-refinement stage of Dai et al.
\cite{dai2026fullsong} from music
to speech, music, and environmental sound with a shared BEST-RQ--Conformer
\cite{chiu2022bestrq,gulati2020conformer}. We retain masked acoustic learning,
mel/chroma reconstruction, and a CTC-based ASR objective, then add clip-level
audio--description alignment and test time-structured and event-description variants.
To isolate their contribution, three paired seeds compare acoustic+CTC,
shuffled-description, and correctly aligned trajectories under matched initialization,
audio exposure and order, control objectives, and optimizer schedule
(Figure~\ref{fig:study-design}). Descriptions come from Music and Sound, leaving Speech
to test transfer without description supervision. Temporal-mean linear probes measure class
structure available through a simple decision surface; a sequence-aware LLM readout
following XARES-LLM \cite{dinkel2026xaresllm} tests whether the frozen sequence remains
useful to a higher-capacity downstream model under the same X-ARES-style tasks and
exposure \cite{zhang2025xares}.

\input{figures/study_design}

Our contributions are twofold. First, we develop a multidomain semantic-refinement
recipe combining masked prediction, spectral/pitch reconstruction, lexical supervision,
and audio--description alignment under uneven annotations. Second, we characterize how
semantic supervision should enter this stack: correct clip-level correspondence
supplies the main transferable semantic signal, while dense acoustic objectives
preserve complementary structure and finer time/event targets add a smaller gain in
linearly accessible semantics. The two readouts reveal where each benefit appears,
and the scaled continuation remains competitive with public encoders.

%% file: figures/study_design.tex
\begin{figure*}[t]
  \centering
  \includegraphics[width=\textwidth]{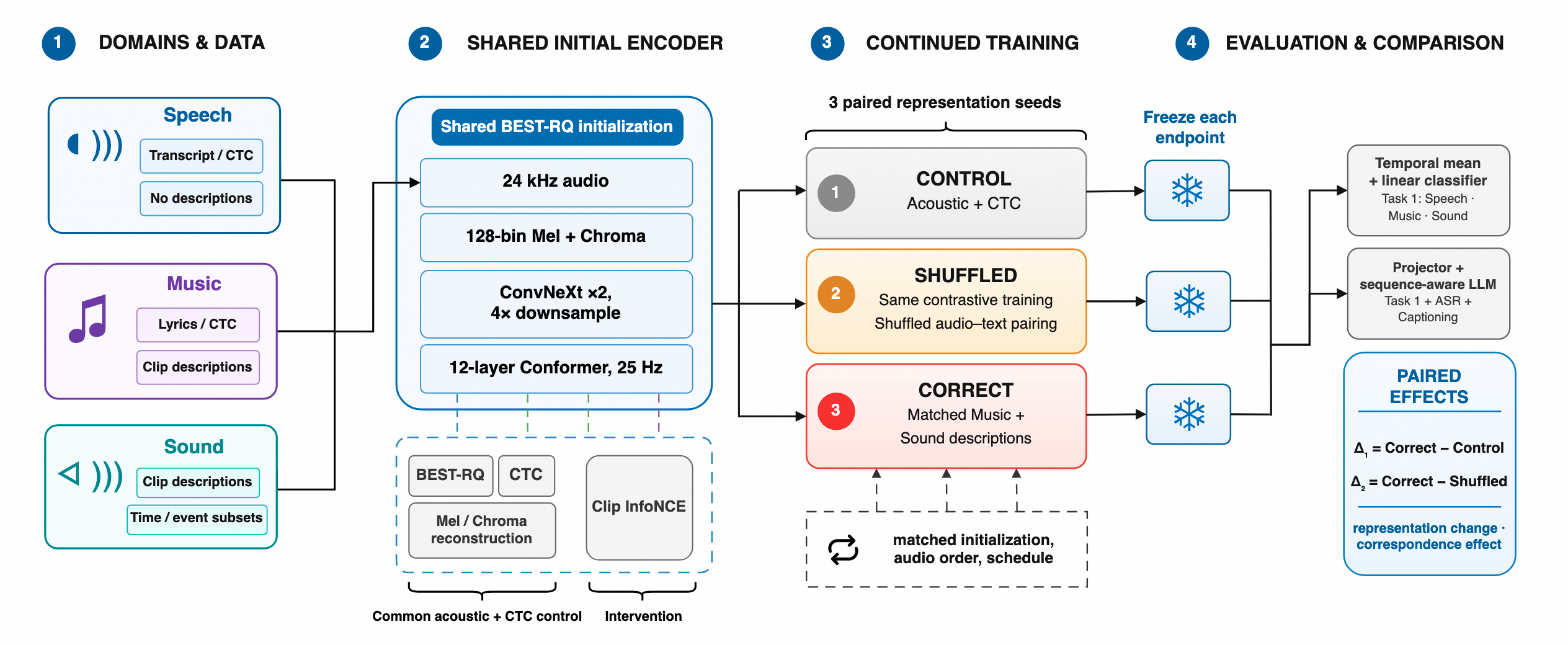}
  \caption{Overview of the matched training intervention and frozen evaluation.
  Speech, Music, and Sound share the BEST-RQ--Conformer initialization but provide
  different supervision. Three paired trajectories produce control, shuffled-description,
  and correctly aligned endpoints, which are frozen and evaluated with identical linear
  and sequence-aware LLM protocols.}
  \label{fig:study-design}
\end{figure*}

%% file: sections/method.tex
\section{Controlled Semantic Refinement}
\label{sec:method}

\subsection{Acoustic and semantic objectives}

We continue a shared BEST-RQ representation across speech, music, and environmental
sound. All trajectories begin from the same 12-block slice of a pretrained 24-block
encoder, which converts
24-kHz audio into 128-bin mel features, reduces the frame rate to 25~Hz with two
ConvNeXt blocks \cite{liu2022convnext}, and processes the sequence with a
1024-dimensional Conformer. Continued BEST-RQ prediction preserves masked acoustic
context. The \emph{acoustic+CTC control} also reconstructs mel and chroma for signal
and pitch structure and applies a CTC-based ASR objective to available transcripts or
lyrics for sequence-aligned lexical cues \cite{graves2006ctc}.

Reconstruction and CTC still leave many global concepts unnamed, especially in music
and environmental sound. We therefore align each clip with its descriptions. The audio
embedding is obtained by masked temporal averaging, projection, and normalization; a
two-layer Transformer over frozen multilingual-BERT token embeddings
\cite{devlin2019bert} produces the text embedding. For normalized embeddings
$\mathbf a_i$ and $\mathbf t_j$, symmetric multi-positive InfoNCE
\cite{elizalde2023clap} is
\begin{align}
 \ell_{a\rightarrow t}
 &=-\frac{1}{|\mathcal A|}\sum_{i\in\mathcal A}
 \log\frac{\sum_{j\in\mathcal P(i)}e^{\gamma\mathbf a_i^\top\mathbf t_j}}
 {\sum_{j\in\mathcal T}e^{\gamma\mathbf a_i^\top\mathbf t_j}}, \nonumber\\[-2pt]
 \mathcal L_{\mathrm{clip}}
 &=\tfrac12(\ell_{a\rightarrow t}+\ell_{t\rightarrow a}).
 \label{eq:clip-infonce}
\end{align}
$\mathcal P(i)$ contains every description view of a recording, including exact
duplicates that would otherwise become false negatives; the reverse term swaps audio
and text anchors. Music and Sound provide descriptions, while Speech continues through
the shared control objectives. Missing targets are masked, and annotated losses are
normalized within domain before fixed domain weighting.

Finer targets reuse the text encoder and contrastive loss. Music time text encodes
ordered section ranges, intensity, and instruments; Sound time text describes
recording-level spatiotemporal dynamics. The time loss applies position-aware query
pooling to the full sequence; the Sound event loss aligns a mean-pooled clip with up
to four event phrases. Missing targets are skipped.

\subsection{Matched training trajectories}

Better downstream scores after adding descriptions could come from correct semantic
pairing or simply from adding another contrastive objective. To separate the two, let
$E_s^{\mathrm{RQ}}$, $E_s^{0}$, $E_s^{\mathrm{shuf}}$, and
$E_s^{\mathrm{clip}}$ denote the RQ-only, acoustic+CTC, shuffled, and correctly paired
frozen encoders for representation seed $s$. For readout $r$ and metric $M_r$, our
primary estimate is
\begin{equation}
  \Delta^{\mathrm{clip-0}}_{s,r}
  =M_r(E_s^{\mathrm{clip}})-M_r(E_s^{0}).
\end{equation}
Within a seed, the three main trajectories share encoder initialization, accepted audio
and order, control objectives, and optimizer schedule. Correct and shuffled training
also share the text-side initialization, caption counts, and contrastive weight;
shuffling only permutes caption bundles within domain and length strata and forbids
self-assignment. Thus correct--control measures the complete alignment component,
whereas correct--shuffled tests the contribution of correct correspondence.

Secondary comparisons place clip alignment in the objective stack. Clip-only and
all-description variants match data, order, control objectives, and clip loss; the
latter adds time/event losses under pooled or domain-normalized aggregation. RQ-only
versus acoustic+CTC tests reconstruction and CTC; removing BEST-RQ, mel, and chroma
tests whether semantics replace or complement dense learning.

%% file: sections/experimental_design.tex
\section{Evaluation Protocol}
\label{sec:design}

\subsection{Training setup and frozen readouts}

We freeze every continued encoder to test whether the information sought by the
training design is already present in its representation. The three training domains
play different roles in this test:
Speech supplies transcripts but no descriptions, Music supplies both lyrics and
descriptions, and Sound supplies descriptions without CTC. Speech therefore reveals
whether description learning transfers beyond the captioned data, while Music and
Sound measure its direct effect. The smaller time/event subsets support the tests of
finer supervision. Table~\ref{tab:objective-routing} summarizes this routing and the
evaluated conditions.

\begin{table}[h]
  \centering
  \caption{Target coverage and matched 50k training conditions. Checkmarks denote
  dense audio-derived objectives; percentages give annotation coverage among
  effective records. ``All desc.'' means clip+time+event; ``domain''/``pool'' specify
  annotated-loss aggregation.}
  \label{tab:objective-routing}
  \vspace{3pt}
  \TableBodySize
  \setlength{\tabcolsep}{0.8pt}
  \begin{tabular*}{\columnwidth}{@{\extracolsep{\fill}}lccccccc@{}}
    \toprule
    Entry & RQ & Mel/Chr. & CTC & Clip & Time & Event & Avg. \\
    \midrule
    \multicolumn{8}{l}{\textit{(a) Target coverage by domain}} \\
    \Speech & \checkmark & \checkmark & 100\% & 0\% & 0\% & 0\% & -- \\
    \Music  & \checkmark & \checkmark & 79.1\% & 100\% & 94.3\% & 0\% & -- \\
    \Sound  & \checkmark & \checkmark & 0\% & 100\% & 37.2\% & 37.3\% & -- \\
    \midrule
    \multicolumn{8}{l}{\textit{(b) Evaluated training conditions}} \\
    RQ-only & \checkmark & -- & -- & -- & -- & -- & -- \\
    Acoustic+CTC & \checkmark & \checkmark & \checkmark & -- & -- & -- & domain \\
    + Shuffled clip & \checkmark & \checkmark & \checkmark & \checkmark & -- & -- & domain \\
    + Correct clip & \checkmark & \checkmark & \checkmark & \checkmark & -- & -- & domain \\
    All desc.; no RQ/Mel/Chr. & -- & -- & \checkmark & \checkmark & \checkmark & \checkmark & domain \\
    All desc.; pool & \checkmark & \checkmark & \checkmark & \checkmark & \checkmark & \checkmark & pool \\
    All desc.; domain & \checkmark & \checkmark & \checkmark & \checkmark & \checkmark & \checkmark & domain \\
    \bottomrule
  \end{tabular*}
\end{table}

Training draws from agreement-governed internal speech, music, and environmental-audio
collections that are separate from the downstream evaluation benchmarks. After
deduplication and validity filtering, we select whole records toward a 5:5:1
Speech/\allowbreak Music/\allowbreak Sound duration target. The 50k controlled runs share the filtered pool,
23.7k accepted audio-hours, batch order, and optimization schedule; accepted hours
measure exposure rather than unique corpus size. Speech/Music text targets are transcripts or lyrics, while
automated internal pipelines provide Music/Sound descriptions. Each controlled
trajectory runs for 50k AdamW updates, with $\beta=.9/.96$, weight decay $10^{-6}$,
gradient clipping at 1, and a 4k-step warmup followed by cosine decay from
$8\!\times\!10^{-5}$ to $8\!\times\!10^{-6}$. RQ/Mel/Chroma have unit weight;
Speech/Music CTC use $.5/.2$, and clip/time/event use $.1/.05/.05$. Routed losses
ramp over 5k steps, and the learnable InfoNCE scale starts at $\gamma_0=1/.07$.

The two readouts answer complementary questions. The \emph{linear probe} mean-pools
valid states and trains one classifier per dataset, asking whether semantic categories
are already accessible through a simple decision surface. The \emph{sequence-aware LLM
readout} follows XARES-LLM \cite{dinkel2026xaresllm}: it retains the frozen sequence and
connects it through a learned projector to rank-8 LoRA-adapted Qwen3-0.6B
\cite{qwen2025qwen3,hu2022lora}, asking whether the same representation remains useful
to a stronger downstream model. Both follow fixed X-ARES-style protocols
\cite{zhang2025xares}, share splits, labels, exposure sequences, and representation
seeds, and hold readout initialization and data order fixed within protocol. Linear
probes train for 6250 steps; the sequence-aware protocol uses 400k downstream examples.

\subsection{Tasks, metrics, and statistical analysis}

We first test whether the representation supports recognition across all three domains.
Task~1 follows X-ARES across 14 datasets. Speech comprises CREMA-D, FSC, LibriCount,
SC-V1, VocalSound, VoxCeleb1, and VoxLingua33; Music comprises FMA, GTZAN, and NSynth;
Sound comprises ESC-50, FSD50K, FSDKaggle2018, and US8K. We use accuracy except mAP
for the FSD datasets, average within domain, and weight the three domain means equally.
This prevents the larger Speech suite from dominating T1.

The sequence-aware LLM also tests whether these representations support generation.
Task~2 ASR averages $1-\mathrm{CER}$ on
AISHELL-1 \cite{bu2017aishell} and $1-\mathrm{WER}$ on LibriSpeech
\cite{panayotov2015librispeech}; captioning uses FENSE \cite{zhou2022fense} on Clotho
\cite{drossos2020clotho} and SongDescriber \cite{manco2023songdescriber}, and DATE on
MECAT \cite{niu2025mecat}. Scores are mapped to a 0--100 higher-is-better scale. We
use their macro-average only as a compact description and interpret ASR and captioning
separately.

Each primary comparison uses three representation-training seeds, except the initial
RQ-only checkpoint. Pairing within seed lets us report mean differences and 95\%
Student-$t$ intervals. For scale context, both readouts also compare the 24-layer
continuation with WavLM Large, MuQ-large, DaSheng-Base, and SPEAR XLarge v2
\cite{chen2022wavlm,zhu2025muq,dinkel2024dasheng,yang2026spear}. These unmatched rows
do not support attribution.

%% file: sections/results.tex
\section{Results}
\label{sec:results}

\subsection{Correct correspondence improves frozen transfer}

The central comparison asks whether descriptions improve the frozen encoder beyond
acoustic and CTC training. Correct alignment raises T1 by
\Result{linearcontrast.p3-p1.equal_domain.mean} points with the linear probe and
\Result{contrast.p3-p1.equal_domain.mean} points with the sequence-aware LLM
(Table~\ref{tab:main-effect}). Every paired seed improves in every domain under both
readouts. The gain is therefore visible after simple temporal averaging and remains
useful when a stronger model reads the full sequence.

\begin{table}[t]
  \centering
  \caption{Task~1 classification under the linear probe and sequence-aware LLM
  readout (\%). Absolute rows are seed means; contrast rows are matched paired
  differences over three representation seeds. Brackets report 95\% paired intervals
  for the T1 contrasts.}
  \label{tab:main-effect}
  \vspace{2pt}
  \TableSize
  \setlength{\tabcolsep}{3.2pt}
  \begin{tabular*}{\columnwidth}{@{\extracolsep{\fill}}lrrrr@{}}
    \toprule
    \multicolumn{5}{l}{\textit{(a) Linear probe}} \\
    Condition & Speech & Music & Sound & T1 avg. \\
    \midrule
    Control & \Result{linear.p1.speech.mean} & \Result{linear.p1.music.mean}
      & \Result{linear.p1.sound.mean} & \Result{linear.p1.equal_domain.mean} \\
    Shuffled & \Result{shuffle.linear.speech.mean} & \Result{shuffle.linear.music.mean}
      & \Result{shuffle.linear.sound.mean} & \Result{shuffle.linear.equal_domain.mean} \\
    Correct & \Best{\Result{linear.p3.speech.mean}} & \Best{\Result{linear.p3.music.mean}}
      & \Best{\Result{linear.p3.sound.mean}} & \Best{\Result{linear.p3.equal_domain.mean}} \\
    \midrule
    Correct $-$ control & +\Result{linearcontrast.p3-p1.speech.mean}
      & +\Result{linearcontrast.p3-p1.music.mean} & +\Result{linearcontrast.p3-p1.sound.mean}
      & \Best{+\Result{linearcontrast.p3-p1.equal_domain.mean}
        [\Result{linearcontrast.p3-p1.equal_domain.ci-low},\Result{linearcontrast.p3-p1.equal_domain.ci-high}]} \\
    Correct $-$ shuffled & +\Result{shuffle.linear.correct-minus-shuffle.speech.mean}
      & +\Result{shuffle.linear.correct-minus-shuffle.music.mean}
      & +\Result{shuffle.linear.correct-minus-shuffle.sound.mean}
      & \Best{+\Result{shuffle.linear.correct-minus-shuffle.equal_domain.mean}
        [\Result{shuffle.linear.correct-minus-shuffle.equal_domain.ci-low},\Result{shuffle.linear.correct-minus-shuffle.equal_domain.ci-high}]} \\
    \bottomrule
  \end{tabular*}

  \vspace{1.5mm}
  \begin{tabular*}{\columnwidth}{@{\extracolsep{\fill}}lrrrr@{}}
    \toprule
    \multicolumn{5}{l}{\textit{(b) Sequence-aware LLM readout}} \\
    Condition & Speech & Music & Sound & T1 avg. \\
    \midrule
    Control & \Result{proxy.p1.speech.mean} & \Result{proxy.p1.music.mean}
      & \Result{proxy.p1.sound.mean} & \Result{proxy.p1.equal_domain.mean} \\
    Shuffled & \Result{shuffle.llm.shuffled.speech.mean}
      & \Result{shuffle.llm.shuffled.music.mean} & \Result{shuffle.llm.shuffled.sound.mean}
      & \Result{shuffle.llm.shuffled.equal_domain.mean} \\
    Correct & \Best{\Result{proxy.p3.speech.mean}} & \Best{\Result{proxy.p3.music.mean}}
      & \Best{\Result{proxy.p3.sound.mean}} & \Best{\Result{proxy.p3.equal_domain.mean}} \\
    \midrule
    Correct $-$ control & +\Result{contrast.p3-p1.speech.mean}
      & +\Result{contrast.p3-p1.music.mean} & +\Result{contrast.p3-p1.sound.mean}
      & \Best{+\Result{contrast.p3-p1.equal_domain.mean}
        [\Result{contrast.p3-p1.equal_domain.ci-low},\Result{contrast.p3-p1.equal_domain.ci-high}]} \\
    Correct $-$ shuffled
      & +\Result{shuffle.llm.correct-minus-shuffle.speech.mean}
      & +\Result{shuffle.llm.correct-minus-shuffle.music.mean}
      & +\Result{shuffle.llm.correct-minus-shuffle.sound.mean}
      & +\Result{shuffle.llm.correct-minus-shuffle.equal_domain.mean}
        [\Result{shuffle.llm.correct-minus-shuffle.equal_domain.ci-low},\Result{shuffle.llm.correct-minus-shuffle.equal_domain.ci-high}] \\
    \bottomrule
  \end{tabular*}
\end{table}

The domain pattern is consistent with where the descriptions enter training. Music and
Sound receive the largest gains, while uncaptioned Speech improves in all six
readout--seed combinations with a smaller effect. Alignment learned from Music and
Sound therefore also strengthens Speech features rather than remaining confined to
the captioned training sources.

The shuffled control separates correct correspondence from simply adding contrastive
training. Shuffled descriptions add only
\Result{shuffle.linear-minus-control.equal_domain.mean} linear T1 points, whereas
correct descriptions add a further
\Result{shuffle.linear.correct-minus-shuffle.equal_domain.mean} points---87\% of the
full control-to-correct gain. Every seed--domain difference favors correct pairing.
The sequence-aware LLM provides complementary evidence: correct pairing exceeds
shuffling by \Result{shuffle.llm.correct-minus-shuffle.equal_domain.mean} T1 points and
gives a consistent +\Result{shuffle.llm.correct-minus-shuffle.caption.mean}-point
captioning advantage. Thus correct correspondence explains most of the linearly
accessible gain; its improvement over the acoustic+CTC control also remains useful to
a full-sequence LLM, with the correspondence-specific benefit clearest in captioning.

\begin{table}[t]
  \centering
  \caption{Detailed three-seed treatment results (\%). Panel (a) gives mean
  Correct $-$ Control Task~1 changes; panel (b) gives sequence-aware LLM Task~2
  means and contrasts.}
  \label{tab:treatment-detail}
  \vspace{2pt}
  \TableSize
  \setlength{\tabcolsep}{3.0pt}
  \begin{tabular*}{\columnwidth}{@{\extracolsep{\fill}}lrr@{}}
    \toprule
    \multicolumn{3}{l}{\textit{(a) Task~1 dataset-level effects}} \\
    Dataset & Linear change & LLM change \\
    \midrule
    \multicolumn{3}{l}{\textit{Speech}} \\
    CREMA-D & +\Result{linearcontrast.p3-p1.task.cremad.mean}
      & +\Result{contrast.p3-p1.task.cremad.mean} \\
    FSC & \Result{linearcontrast.p3-p1.task.fsc.mean}
      & +\Result{contrast.p3-p1.task.fsc.mean} \\
    LibriCount & +\Result{linearcontrast.p3-p1.task.libricount.mean}
      & +\Result{contrast.p3-p1.task.libricount.mean} \\
    SC-V1 & +\Result{linearcontrast.p3-p1.task.scv1.mean}
      & +\Result{contrast.p3-p1.task.scv1.mean} \\
    VocalSound & +\Result{linearcontrast.p3-p1.task.vocalsound.mean}
      & +\Result{contrast.p3-p1.task.vocalsound.mean} \\
    VoxCeleb1 & +\Result{linearcontrast.p3-p1.task.voxceleb1.mean}
      & +\Result{contrast.p3-p1.task.voxceleb1.mean} \\
    VoxLingua33 & \Result{linearcontrast.p3-p1.task.voxlingua33.mean}
      & +\Result{contrast.p3-p1.task.voxlingua33.mean} \\
    \multicolumn{3}{l}{\textit{Music}} \\
    FMA & +\Result{linearcontrast.p3-p1.task.fma.mean}
      & +\Result{contrast.p3-p1.task.fma.mean} \\
    GTZAN & +\Result{linearcontrast.p3-p1.task.gtzan.mean}
      & +\Result{contrast.p3-p1.task.gtzan.mean} \\
    NSynth & +\Result{linearcontrast.p3-p1.task.nsynth.mean}
      & \Result{contrast.p3-p1.task.nsynth.mean} \\
    \multicolumn{3}{l}{\textit{Sound}} \\
    ESC-50 & +\Result{linearcontrast.p3-p1.task.esc50.mean}
      & +\Result{contrast.p3-p1.task.esc50.mean} \\
    FSD50K & +\Result{linearcontrast.p3-p1.task.fsd50k.mean}
      & +\Result{contrast.p3-p1.task.fsd50k.mean} \\
    FSDKaggle2018 & +\Result{linearcontrast.p3-p1.task.fsdkaggle2018.mean}
      & +\Result{contrast.p3-p1.task.fsdkaggle2018.mean} \\
    UrbanSound8K & +\Result{linearcontrast.p3-p1.task.urbansound8k.mean}
      & +\Result{contrast.p3-p1.task.urbansound8k.mean} \\
    \bottomrule
  \end{tabular*}

  \vspace{1.5mm}
  \begin{tabular*}{\columnwidth}{@{\extracolsep{\fill}}lrrr@{}}
    \toprule
    \multicolumn{4}{l}{\textit{(b) Sequence-aware LLM Task~2}} \\
    Condition / contrast & ASR & Caption & Task~2 \\
    \midrule
    Control & \Result{shuffle.llm.control.asr.mean}
      & \Result{shuffle.llm.control.caption.mean}
      & \Result{shuffle.llm.control.task2.mean} \\
    Shuffled & \Result{shuffle.llm.shuffled.asr.mean}
      & \Result{shuffle.llm.shuffled.caption.mean}
      & \Result{shuffle.llm.shuffled.task2.mean} \\
    Correct & \Best{\Result{shuffle.llm.correct.asr.mean}}
      & \Best{\Result{shuffle.llm.correct.caption.mean}}
      & \Best{\Result{shuffle.llm.correct.task2.mean}} \\
    \midrule
    Correct $-$ control & +\Result{contrast.p3-p1.asr.mean}
      & +\Result{contrast.p3-p1.caption.mean}
      & +\Result{contrast.p3-p1.task2.mean} \\
    Correct $-$ shuffled & +\Result{shuffle.llm.correct-minus-shuffle.asr.mean}
      & +\Result{shuffle.llm.correct-minus-shuffle.caption.mean}
      & +\Result{shuffle.llm.correct-minus-shuffle.task2.mean} \\
    \bottomrule
  \end{tabular*}
\end{table}

The task-level results in Table~\ref{tab:treatment-detail} support the same picture.
Correct alignment improves 12 of 14 datasets with the linear probe and 13 with the LLM;
FSC, VoxLingua33, and NSynth account for the readout-dependent signs. Task~2 further
separates the effect: correct pairing exceeds shuffling by
\Result{shuffle.llm.correct-minus-shuffle.caption.mean} captioning points but only
\Result{shuffle.llm.correct-minus-shuffle.asr.mean} ASR points. Description alignment
therefore changes broad semantic transfer more clearly than ASR.

\subsection{How clip alignment fits the objective stack}

The main comparison establishes the value of clip alignment; Table~\ref{tab:secondary}
places it in the full objective stack through matched contrasts. In the single-seed
diagnostic, Acoustic+CTC improves linear/LLM T1 by
\Result{linear.p1seed1342-minus-p0rq.equal_domain}/
\Result{proxy.p1seed1342-minus-p0rq.equal_domain} points and raises ASR from
\Result{proxy.p0rq.asr.mean} to \Result{proxy.p1seed1342.asr}, while captioning remains
unchanged. The linear gain is concentrated in Speech
(+\Result{linear.p1seed1342-minus-p0rq.speech}); Music and Sound change by
\Result{linear.p1seed1342-minus-p0rq.music} and
\Result{linear.p1seed1342-minus-p0rq.sound} points. This diagnostic indicates that the
reconstruction-and-CTC bundle primarily strengthens recognition; it does not attribute
the change to CTC alone.

The next comparison shows why the semantic objective should not replace acoustic
learning. Restoring BEST-RQ, mel, and chroma to the semantic system improves T1 by
\Result{p5-p4.linear.equal_domain.mean} points
[\Result{p5-p4.linear.equal_domain.ci-low},
\Result{p5-p4.linear.equal_domain.ci-high}] under the linear probe and
\Result{p5-p4.llm.equal_domain.mean} points
[\Result{p5-p4.llm.equal_domain.ci-low},
\Result{p5-p4.llm.equal_domain.ci-high}] under the LLM. Sound gains
\Result{p5-p4.linear.sound.mean}/\Result{p5-p4.llm.sound.mean} points, and captioning
also rises by \Result{p5-p4.llm.caption.mean}. The agreement across both readouts
supports a combined representation: descriptions add semantics while dense objectives
retain useful acoustic structure.

Finer supervision provides a further semantic gain, most clearly in the simple
readout. Time/event targets raise linear T1 by
\Result{linearcontrast.p5-p3.equal_domain.mean} points
([\Result{linearcontrast.p5-p3.equal_domain.ci-low},
\Result{linearcontrast.p5-p3.equal_domain.ci-high}]), indicating that the additional
targets make semantic structure more accessible to a linear classifier. LLM T1 also
moves upward by \Result{contrast.p5-p3.equal_domain.mean} points, while Task~2 shows
only small changes, so the benefit is concentrated in classification rather than generation.
Domain-wise rather than pooled averaging changes linear/LLM T1 by
\Result{linearcontrast.p5-p2.equal_domain.mean}/+
\Result{contrast.p5-p2.equal_domain.mean} points, leaving the two aggregation recipes
with similar aggregate transfer.

\begin{table}[t]
  \centering
  \caption{Objective-stack and source analyses (\%). Panel (a) reports the named
  condition minus its reference; \textsuperscript{1} marks a single-seed matched
  diagnostic, while the remaining rows use three paired seeds. Panel (b) is a
  single-seed source intervention. Mu./So.: Music/Sound macros.}
  \label{tab:secondary}
  \vspace{2pt}
  \TableSize
  \setlength{\tabcolsep}{2.5pt}
  \begin{tabular*}{\columnwidth}{@{\extracolsep{\fill}}lrrrr@{}}
    \toprule
    \multicolumn{5}{l}{\textit{(a) Matched objective-stack contrasts}} \\
    Contrast & Lin. T1 & LLM T1 & ASR & Cap. \\
    \midrule
    Ac.+CTC vs. RQ-only\textsuperscript{1}
      & +\Result{linear.p1seed1342-minus-p0rq.equal_domain}
      & +\Result{proxy.p1seed1342-minus-p0rq.equal_domain}
      & +\Result{proxy.p1seed1342-minus-p0rq.asr}
      & \Result{proxy.p1seed1342-minus-p0rq.caption} \\
    Dense $-$ no dense
      & +\Result{p5-p4.linear.equal_domain.mean}
      & +\Result{p5-p4.llm.equal_domain.mean}
      & \Result{p5-p4.llm.asr.mean}
      & +\Result{p5-p4.llm.caption.mean} \\
    Time/event $-$ clip
      & +\Result{linearcontrast.p5-p3.equal_domain.mean}
      & +\Result{contrast.p5-p3.equal_domain.mean}
      & \Result{contrast.p5-p3.asr.mean}
      & +\Result{contrast.p5-p3.caption.mean} \\
    Domain $-$ pool avg.
      & \Result{linearcontrast.p5-p2.equal_domain.mean}
      & +\Result{contrast.p5-p2.equal_domain.mean}
      & \Result{contrast.p5-p2.asr.mean}
      & +\Result{contrast.p5-p2.caption.mean} \\
    \bottomrule
  \end{tabular*}

  \vspace{2mm}
  \setlength{\tabcolsep}{2.0pt}
  \begin{tabular*}{\columnwidth}{@{\extracolsep{\fill}}lrrrrrr@{}}
    \toprule
    \multicolumn{7}{l}{\textit{(b) Exploratory description-source intervention}} \\
      & \multicolumn{3}{c}{Linear} & \multicolumn{3}{c}{Sequence-aware LLM} \\
    Source & Mu. & So. & T1 & Mu. & So. & T1 \\
    \midrule
    None & \Result{linearfactorial.none.music} & \Result{linearfactorial.none.sound}
      & \Result{linearfactorial.none.equal_domain} & \Result{factorial.none.music}
      & \Result{factorial.none.sound} & \Result{factorial.none.equal_domain} \\
    Music & \Best{\Result{linearfactorial.music.music}} & \Result{linearfactorial.music.sound}
      & \Result{linearfactorial.music.equal_domain} & \Best{\Result{factorial.music.music}}
      & \Result{factorial.music.sound} & \Result{factorial.music.equal_domain} \\
    Sound & \Result{linearfactorial.sound.music} & \Result{linearfactorial.sound.sound}
      & \Result{linearfactorial.sound.equal_domain} & \Result{factorial.sound.music}
      & \Best{\Result{factorial.sound.sound}} & \Result{factorial.sound.equal_domain} \\
    Music + Sound & \Result{linearfactorial.both.music} & \Best{\Result{linearfactorial.both.sound}}
      & \Best{\Result{linearfactorial.both.equal_domain}} & \Result{factorial.both.music}
      & \Result{factorial.both.sound} & \Best{\Result{factorial.both.equal_domain}} \\
    \bottomrule
  \end{tabular*}
\end{table}

Panel (b) offers a final, exploratory view of where the semantic gain originates. Each
single description source gives the largest point estimate in its own domain under at
least one readout. Combining Music and Sound produces the broadest profile and the
highest overall T1, favoring both sources when domain-balanced transfer is the goal.

\subsection{Scaling and public-model context}

\begin{table}[t]
  \centering
  \caption{Single-endpoint public comparison (\%; higher is better). ASR/Caption
  use the LLM readout. Rows differ in architecture, capacity, data, and optimization;
  our 24L/105k row uses its preselected layer-12 representation.}
  \label{tab:public-context}
  \vspace{2pt}
  \TableSize
  \setlength{\tabcolsep}{3.0pt}
  \begin{tabular*}{\columnwidth}{@{\extracolsep{\fill}}lrrrr@{}}
    \toprule
    Encoder & Lin. T1 & LLM T1 & ASR & Caption \\
    \midrule
    WavLM Large & \Result{linear.public.wavlm.equal_domain}
      & \Result{xares.public.wavlm.equal_domain} & \Result{xares.public.wavlm.asr}
      & \Result{xares.public.wavlm.caption} \\
    MuQ-large & \Result{linear.public.muq.equal_domain}
      & \Result{xares.public.muq.equal_domain} & \Result{xares.public.muq.asr}
      & \Result{xares.public.muq.caption} \\
    DaSheng-Base & \Result{linear.public.dasheng.equal_domain}
      & \Best{\Result{xares.public.dasheng.equal_domain}} & \Result{xares.public.dasheng.asr}
      & \Result{xares.public.dasheng.caption} \\
    SPEAR XLarge v2 & \Best{\Result{linear.public.spear.equal_domain}}
      & \Result{xares.public.spear.equal_domain} & \Result{xares.public.spear.asr}
      & \Best{\Result{xares.public.spear.caption}} \\
    Ours, 24L/105k & \Result{linear.full.domainsemantic105k.equal_domain}
      & \Result{xares.full.domainsemantic105k.equal_domain}
      & \Best{\Result{xares.full.domainsemantic105k.asr}}
      & \Result{xares.full.domainsemantic105k.caption} \\
    \bottomrule
  \end{tabular*}
\end{table}

The full 24-layer encoder reaches
\Result{linear.full.domainsemantic105k.equal_domain} linear T1 and
\Result{xares.full.domainsemantic105k.equal_domain} LLM T1. It trails SPEAR and
DaSheng, exceeds WavLM and MuQ under both readouts, and has the highest ASR point
estimate in Table~\ref{tab:public-context}. It follows the same routing policy and accumulates 99.7k accepted
audio-hours, an exposure count rather than unique corpus size or total pretraining.
This unmatched comparison supports the recipe at scale; the matched 12-block study
remains the basis for attribution.

%% file: sections/conclusion.tex
\section{Conclusion}
\label{sec:conclusion}

Correct audio--description alignment improves all three domains under linear and
sequence-aware LLM readouts; shuffling explains little of the linear gain.
BEST-RQ, spectral/pitch reconstruction, and CTC remain complementary, while finer
targets mainly improve classification. These results favor refining an
acoustic--lexical foundation with correctly paired descriptions. The competitive
24-layer run supports the recipe beyond the controlled study.